\documentstyle[sprocl,psfig]{article}

\input{psfig}
\def\be{\begin{equation}}
\def\ee{\end{equation}}
\def\bea{\begin{eqnarray}}
\def\eea{\end{eqnarray}}
\def\lsim{\compoundrel<\over\sim}
\def\compoundrel#1\over#2{\mathpalette\compoundreL{{#1}\over{#2}}}
\def\compoundreL#1#2{\compoundREL#1#2}
\def\compoundREL#1#2\over#3{\mathrel
     {\vcenter{\hbox{$\buildrel{#1#2}\over{#1#3}$}}}}

\begin{document}

\title{{\em Ab initio} MOLECULAR DYNAMICS STUDY OF D$_2$ DESORPTION FROM
Si\,(100)}

\author{Axel Gross, Michel Bockstedte, and Matthias Scheffler}

\address{Fritz-Haber-Institut, Faradayweg 4-6,
D-14195~Berlin-Dahlem, Germany}


\maketitle\abstracts{{\it Ab initio} molecular dynamics calculations 
of deuterium desorbing from Si(100) have been performed in order 
to monitor the energy redistribution among the hydrogen and silicon degrees of
freedom during the desorption process. The calculations show that
part of the potential energy at the transition
state to desorption is transferred to the silicon lattice.
The deuterium molecules leave the surface vibrationally hot and
rotationally cold, in agreement with experiments; the mean kinetic energy,
however, is larger than found in experiments.
}

\section{Introduction}
Hydrogen adsorption on and desorption from Si surfaces are of great
technological relevance for, e.g., the etching and passivation of
Si surfaces or the growth of Si crystals
(see, e.g., Ref.~[1] and references therein). 
Besides, the dynamics of the hydrogen interaction with Si surfaces is 
also of fundamental interest  
due, among others, to the so-called barrier puzzle:
While the sticking coefficient of molecular hydrogen on Si surfaces
is very small \cite{Lie90,Kol94b,Bra95}
indicating a high barrier to adsorption, in desorption experiments
an almost thermal energy distribution of the molecules was found 
\cite{Kol94a} indicating a low barrier to adsorption. In order to 
explain this puzzle it was suggested to take the strong surface rearrangement 
of Si upon hydrogen adsorption into account: \cite{Kol94b,Bre94} 
The hydrogen molecules impinging on
the Si substrate from the gas phase encounter a Si configuration
which is unfavorable for dissociation, while desorbing hydrogen molecules
leave the surface from a rearranged Si configuration with a low barrier
in the hydrogen coordinates.
The lattice rearrangement energy was assumed to be about 0.7~eV,\cite{Bre94}
a value in reasonable agreement with the transition state energy obtained
by cluster calculations.\cite{Rad96,Nac94,Jin95}
Density-functional calculations of the potential energy surface (PES)
of H$_2$/Si\,(100) using the supercell approach \cite{Kra94,Peh95} 
provided a detailed microscopic description of the interaction. 
The transition state energy was determined to be only
$\sim$ 0.3~-~0.4~eV with the substrate rearrangement energy being
merely $\lsim$~50{\%} of this value.

The calculation of PESs is an important prerequisite for
understanding reaction dynamics. For a quantitative analysis, 
however, a calculation of the dynamics is indispensable.
We have therefore performed {\it ab initio} molecular dynamics
calculations to monitor the energy distribution of D$_2$ molecules
desorbing from Si\,(100). We will show that part of the potential 
energy at the barrier position is indeed transferred to
the silicon lattice, the mean kinetic energy of desorbing molecules,
however, is larger than found in experiments.

\section{Calculational details}
In our {\it ab initio} molecular dynamics calculations 
\cite{Bock96} the forces
necessary to integrate the classical equations of motion are
determined by density-functional calculations. The exchange-correlation
functional is treated in the generalized gradient approximation 
(GGA) \cite{Per92}. 
In previous slab studies the total energies were calculated within
the local density approximation (LDA) with {\it a posteriori} GGA
corrections~\cite{Peh95}. The main effects of using the GGA 
in the complete self-consistent cycle
are a small increase of the theoretical lattice constant of Si \cite{Mol95} 
and a slight rise in the barrier height from $E_b = 0.3$~eV \cite{Peh95} to
$E_b =0.4$~eV. To correctly represent the up and down buckling of the
clean Si(100) surface we use a (2$\times$2) surface unit cell. 
The Si slab consists of five atomic layers.
The topmost three of them are free to move in the
molecular dynamics simulations, while the remaining two layers are
fixed at their bulk positions. The density-functional calculations
are performed with two {\bf k}-points in the irreducible part of the
Brillouin zone and 40~Ry cutoff energy.
The equations of motion are numerically integrated within a
predictor-corrector scheme with a time step of 1.2~fs. 
The calculations have been performed on 
typically 64 nodes of the Cray T3D of the Konrad-Zuse-Zentrum, Berlin.

\section{Results}

\begin{figure}[t]
\unitlength1cm
\begin{center}
   \begin{picture}(10,4.0)
      \includegraphics{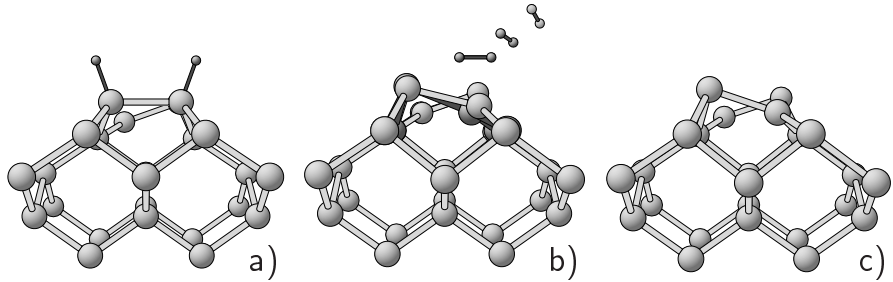}
   \end{picture}
\end{center}
\caption{a) Hydrogen covered Si\,(100) surface (monohydride).
b) Snapshots of a trajectory of D$_2$ desorbing from Si\,(100)
starting at the trasnsition state with the Si atoms initial at rest. 
The dark Si atoms correspond to the
Si positions after the desorption event. 
c) Clean anti-buckled Si\,(100) surface.
\label{figdes}}
\end{figure}

Since the barrier to associative desorption
of hydrogen from Si\,(100) is rather high ($E_d = 2.5$~eV),\cite{Peh95}
there is no sense in performing molecular dynamics calculations starting 
with the deuterium atoms at the adsorption sites because of the extremely
low number of desorption events.
Therefore we started the desorption trajectories
close to the transition state for dissociative adsorption
which was determined in the earlier study.\cite{Peh95}
In total we have computed 40 trajectories of D$_2$ desorbing
from Si\,(100). Eight trajectories were determined with the
Si lattice initially at rest, i.e. at a surface temperature of $T_s = 0$~K,
in order to specifically monitor desorption trajectories starting
at the transition state. 
Figure~\ref{figdes} illustrates the strong surface rearrangement of Si\,(100)
upon hydrogen adsorption/desorption. 
Snapshots of a calculated trajectory are shown in Fig.~\ref{figdes}b).
The dark Si atoms
correspond to the relaxation of the Si lattice after the desorption event.
Approximately 0.1~eV of the potential energy at the transition state
is transferred to vibrations of the Si lattice which is a rather large
amount compared to hydrogen/metal systems.

The desorption experiments determining the mean kinetic energy of
D$_2$ desorbing from Si\,(100) were done with a rather 
high surface temperature of
$T_s \approx 920$~K. In order to simulate these experimental conditions,
we have performed {\it ab initio} molecular dynamics calculations
with initial conditions corresponding to the experimental surface
temperature. The system was allowed to equilibrate for more than 500~fs, 
whereby the deuterium atoms were kept close to the transition
state by an additional cage-like potential. The additional potential
was then switched off and the energy distribution of the D$_2$ molecules
desorbing from the thermal surface were monitored. In total 32
``thermal'' desorption trajectories were calculated. The mean total,
kinetic, vibrational, and rotational energies
of the desorbing molecules are listed in Table~\ref{tabdes} (note
that $k_B T_s = 0.079$~eV). 

The results show vibrational heating, 
i.e. $\langle E_{vib} \rangle > k_B T_s$, and rotational cooling, 
i.e. $\langle E_{rot} \rangle < k_B T_s$, in agreement with the
experiment.\cite{Kol92} The mean kinetic energy, however, is much 
larger than the experimental value of 
$\langle E_{kin} \rangle^{exp} = 0.165$~eV. The difference between
the experimental and theoretical results corresponds roughly to the
barrier height $E_b$. A closer analysis of the trajectories \cite{Gro96} 
reveals that still
approximately 0.1~eV of the potential energy at the transition state is 
transferred to the Si lattice, however, due to the Si lattice vibrations
the mean adsorption barrier is increased by roughly the same amount. 
Possible contributions to the discrepancy between theory and experiment
could be: 
(i) insufficient statistics, i.e., too few trajectories computed, 
(ii)~quantum mechanical effects (e.g., tunneling and zero-point effects)
not taken into account in the classical molecular dynamics, 
(iii) dissipation channels not considered, e.g., electronic excitations, 
(iv) limitations of the GGA functional, 
(v) experimental uncertainties. 
Certainly there is a strong need for future theoretical and experimental
studies of the hydrogen on silicon system.

\begin{table}[t]
\caption{Mean energy distribution averaged over 32 trajectories of D$_2$ 
molecules desorbing from a Si(100) surface at a surface temperature 
of $T_s = 920$~K.
\label{tabdes}}
\vspace{0.4cm}
\begin{center}
\begin{tabular}{|c|c|c|c|}
\hline
$\langle E_{tot} \rangle$ & $0.72  \pm 0.17$ eV &
$\langle E_{vib} \rangle$ & $0.11  \pm 0.09$ eV \\ 
$\langle E_{kin} \rangle$ & $0.58  \pm 0.13$ eV & 
$\langle E_{rot} \rangle$ & $0.03  \pm 0.05$ eV \\  
\hline
\end{tabular}
\end{center}
\end{table}

\mbox{ }

\vspace{-1.3cm}

\mbox{ }

\section*{Acknowledgments}
We gratefully acknowledge a grant of the Konrad-Zuse-Zentrum, Berlin,
for using its computational resources. We thank E. Pehlke, P. Kratzer, 
W. Brenig, and E. Hasselbrink for helpful discussions.

\end{document}